\documentstyle[12pt]{article}
\catcode `\@=11
\def\numberbysection{\@addtoreset{equation}{section}
   \renewcommand{\theequation}{\thesection.\arabic{equation}}}
\newcommand{\be}{\begin{equation}}
\newcommand{\ee}{\end{equation}}
\newcommand{\bea}{\begin{eqnarray}}
\newcommand{\eea}{\end{eqnarray}}
\newcommand{\rar}{\rightarrow}

\newcommand{\vac}{\emptyset}
\newcommand{\qin}{q^{-1}}
\newcommand{\si}{\sigma}
\newcommand{\al}{\alpha}
\newcommand{\bt}{\beta}
\newcommand{\g}{\Gamma}
\newcommand{\ket}[1]{\left| #1\right\rangle}    %ket
\newcommand{\build}[3]{\mathrel{\mathop{#1}
\limits_{#2}^{#3} }}
\numberbysection
\begin{document}
\begin{titlepage}
\baselineskip 0.3in
\null
\begin{center}
\large
{\bf Reaction-Diffusion Processes}\\
{\bf described by}\\
{\bf Three-State Quantum Chains and Integrability}
\vskip 0.5in
\normalsize
{Silvio R. Dahmen\footnotemark[1]
\footnotetext[1]{e-mail:dahmen@pib1.physik.uni-bonn.de}
}
\vskip 0.2in
Physikalisches Institut der Universit\"at Bonn \\
Nu{\ss}allee 12\\
53115 Bonn Germany
\vskip 0.9in
\end{center}
%
%abstract
%
\begin{abstract}
\hspace{\parindent}
The master equation of one-dimensional three-species
reaction-diffusion processes is mapped onto an imaginary-time
Schr\"odinger equation. In many cases the Hamiltonian obtained
is that of an integrable quantum chain.
Within this approach we search for all $3$-state integrable
quantum chains whose spectra are known and which are related
to diffusive-reactive systems.
Two integrable models are found to appear
naturally in this context: the
$U_{q}\widehat{SU(2)}$-invariant model with
external fields and
the $3$-state $U_q SU(P/M)$-invariant Perk-Schultz
models with external fields. A nonlocal similarity
transformation which brings the Hamiltonian governing
the chemical processes to the known standard forms is
described, leading in the case of periodic boundary conditions
to a generalization of the Dzialoshinsky-Moriya interaction.
\end{abstract}
\vskip 0.2in
\begin{flushleft}
\parbox[t]{3.5cm}{\bf Key words:}
\parbox[t]{10cm}{Reaction-diffusion models, integrable
                   quantum chains, Bethe Ansatz}
\\[2mm]
\parbox[t]{3.5cm}{\bf PACS numbers:}
\parbox[t]{10cm}{05.50, 64.60, 75.10J}
\\[2mm]
May 1994
\\[2mm]
cond-mat/9405031
\end{flushleft}
\end{titlepage}
\newpage
\setcounter{page}{1}
\section{Introduction}
\hspace{\parindent}
Since the pioneering work of Smoluchowski in 1917 \cite{smo},
reaction-diffusion-limited processes have had a forefront
position in nonequilibrium statistical Physics.
They can be portrayed as bimolecular
processes of the type $A+B \build {\rightleftharpoons}{r}{k} C+D$
where molecules of species $A$ and $B$ ($C$ and $D$)
or two different states of the same molecule react to form
$C$ and $D$ ($A$ and $B$) with a reaction rate $k$ ($r$).
Particularly in the last decade a
great amount of research has been reported (see
\cite{bramson}-\cite{avra} and references therein).
The interest has been centered mainly on
irreversible ($r=0$) and vacuum-driven chemical reactions,
i.e. those for which at least
one of the final products is an inert state
(a precipitate or a non-reacting molecule)
denoted by $\vac$.
The commonly studied reactions are
\vskip 0.2in
\noindent
{\bf $1$. Diffusion}
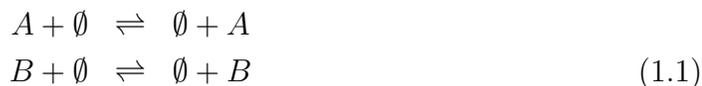
\bea\label{eq:diff}
A+\vac&\rightleftharpoons&\vac+A\nonumber\\
B+\vac&\rightleftharpoons&\vac+B
\eea
{\bf $2$. Interchange}
\bea
A+B&\rightleftharpoons&B+A
\eea
{\bf $3$. Death}
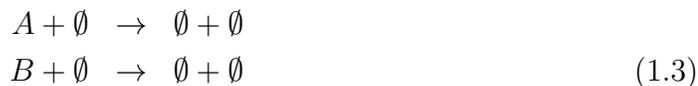
\bea
A+\vac&\rar&\vac+\vac\nonumber\\
B+\vac&\rar&\vac+\vac
\eea
{\bf $4$. One- and Two-Species Annihilation
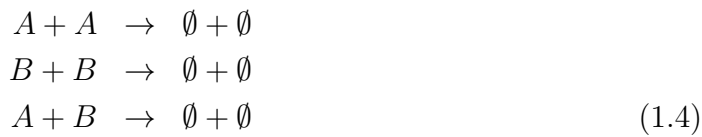
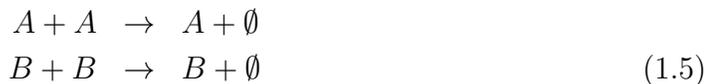
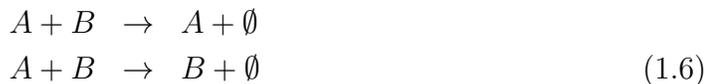
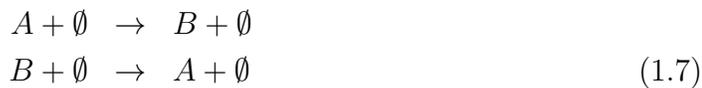
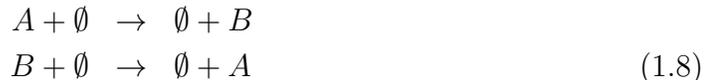
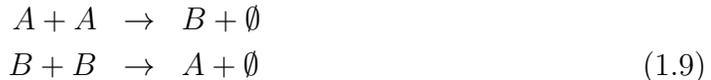
\cite{priv}}
\bea
A+A&\rar&\vac+\vac\nonumber\\
B+B&\rar&\vac+\vac\nonumber\\
A+B&\rar&\vac+\vac
\eea
{\bf $5$. Coagulation \cite{red}}
\bea
A+A&\rar&A+\vac\nonumber\\
B+B&\rar&B+\vac
\eea
{\bf $6$. Trapping \cite{chop}}
\bea
A+B&\rar&A+\vac\nonumber\\
A+B&\rar&B+\vac
\eea
{\bf $7$. Mutation}
\bea
A+\vac&\rar&B+\vac\nonumber\\
B+\vac&\rar&A+\vac
\eea
{\bf $8$. Transmutation}
\bea
A+\vac&\rar&\vac +B\nonumber\\
B+\vac&\rar&\vac +A
\eea
{\bf $9$. Polymerisation \cite{spouge}}
\bea\label{eq:pol}
A+A&\rar&B+\vac\nonumber\\
B+B&\rar&A+\vac
\eea

The apparent simplicity of the processes depicted
above is quite deceiving.
{}From the great variety of nonequilibrium problems,
reaction-diffusion
processes are one of the most difficult to tackle.
An important step towards circumventing these difficulties,
largely of a mathematical nature, was taken
by Glauber in 1963 \cite{glauber} and
subsequently explored by many authors \cite{heims}.
In a pursuit to
understand nonequibrium systems in terms
of the more treatable equilibrium ones, he devised an ingenious
way of using classical spin systems to study
the problem of critical dynamics
by means of a master equation approach. This opened the
possibility of employing results from spin chains
in the context of nonequilibrium problems.
The status quo remained however
pratically unaltered. This was so since the developments
achieved on the theory of spin chains
(or more generally speaking quantum chains), which ultimately
lead us into the concept of integrability \cite{devega},
were made quite independently and the gap remained.
Only recently it was realized
\cite{kandel} that a class of problems regarding
shrinking domains of Ising spins could be
understood in terms of the integrable six-vertex
model in one of its critical manifolds \cite{baxter}.
Besides, using a master equation approach,
a larger class of nonhermitian and integrable $q$-deformed models
were shown to appear naturally as the time evolution operators of
several reaction-diffusion processes \cite{alc3}.

Within this spirit, our aim in this paper is to find all three-state
integrable quantum chains whose Hamiltonians are time-evolution
operators of diffusion-reaction processes and whose spectra are
known or can be easily calculated.
Our motivation is twofold: the equivalence
of the spectra guarantees us the equivalence of the phase
diagram and the physical behavior of
chemical systems is then in principle determined.
Consider the long-time behavior of systems governed by the
vacuum-driven reactions given above, as an example.
The systems will relax to a final state where there are no
particles left, the mean concentration
for particle $A$ (or $B$) decaying as
\be
c_{A,B}\sim\left\{ \begin{array}{c}
t^{-\al}\\e^{-t/\tau }
\end{array}
\right.
\ee
where $\al$ and $\tau$ are characteristic of each problem.
Borrowing the field-theoretic jargon we can talk of a massless
and massive phase if by these we mean the behavior depicted
above. In the one-dimensional annihilation and coagulation models
one has $\al = \frac{1}{2}$ which imply in a
slower decay towards the vacuum state. The inclusion of certain
kinds of processes (or reversible reactions) leads to a local
steady state, thus driving the system towards a quicker
(exponential) decay-regime. We shall see in the next sections
that these behaviors correspond, in the quantum chains to which
these models are related, to the massless and massive regimes
respectively.

Our second motivation comes from the fact that the
connection with integrable systems provides
us with methods of calculating
physical quantities exactly.
In this respect, we were particularly motivated by
the work of Gwa and Spohn \cite{gwa} who calculated
the dynamical scaling exponent of the discrete
Noisy Burgers Equation. This equation was introduced
by Burgers \cite{burgers} as a model for
turbulent flow. The discretization leads to
an asymmetric two-state diffusion problem described by the
reaction $A+\vac \build {\rightleftharpoons}{r}{k} \vac +A$
where the rates $k$ and $r$ are different.
The dynamics of this problem is governed by the Hamiltonian
of the six-vertex model.

We summarize the results of this paper in what follows.
We found that
a large class of chemical reactions composed
of several simultaneous processes can
be understood in terms of two quantum chains, namely the
$U_{q}\widehat{SU(2)}$-invariant model and the $3$-state
$U_{q}SU(P/M)$-invariant Perk-Schultz models. Using the standard
basis of matrices $(E^{kl})_{a,b}=\delta_{k,a}\delta_{l,b}$ the
first chain reads \cite{alc2}
\bea\label{eq:ohio}
H^{\prime}&=&H_{0}^{\prime}+H_{1}^{\prime}\nonumber\\
H_{0}^{\prime}&=&-\sum_{i=1}^{L-1}\left( E_{i}^{01}E_{i+1}^{10}+
E_{i}^{10}E_{i+1}^{01}+E_{i}^{02}E_{i+1}^{20}+E_{i}^{20}E_{i+1}^{02}
\right) +v \varepsilon_{i}^{0} \varepsilon_{i+1}^{0}\nonumber\\
&&+w\left( \varepsilon_{i}^{0}+\varepsilon_{i+1}^{0}\right)
+a\left( \varepsilon_{i}^{0}-\varepsilon_{i+1}^{0} \right)
\nonumber\\
H_{1}^{\prime}&=&-g\sum_{i=1}^{L} \varepsilon_{i}^{z}\nonumber\\
\varepsilon^{0}&=&E^{11}+E^{22} \qquad \varepsilon^{z}
=E^{11}-E^{22}\nonumber\\
\eea
where $H^{\prime}_{0}$ is $U_{q}\widehat{SU(2)}$-invariant
and the symmetry-breaking $H_{1}^{\prime}$, which commutes with
$H_{0}^{\prime}$, acts as an external field and does not spoil
the integrability of the model.
$H_{0}^{\prime}$ has the same spectrum
(apart from degeneracies) as that of the
spin-$\frac{1}{2}$ Heisenberg model with an external field
and a surface term
\be\label{eq:heisen}
H^{XXZ}=-\frac{1}{2} \sum_{i}^{L-1}\biggl\{ \si_{i}^{x}
\si_{i+1}^{x} +\si_{i}^{y}\si_{i+1}^{y}+\Delta \si_{i}^{z}
\si_{i+1}^{z} +h\left( \si_{i}^{z}+\si_{i+1}^{z}\right)
+a \left(\si_{i}^{z}-\si_{i+1}^{z}\right) +\beta\biggr\}
\ee
The phase diagram is well known \cite{mc}:
without external field, the model is massless and conformal invariant
when $-1\leq \Delta \leq 1$, massive with a ferromagnetic ground
state for $\Delta >1$ and massive with an antiferromagnetic ground state
for $\Delta < -1$. When $h \ne 0$
the system is massive commensurate for $\Delta > 1-h$ and
massless incommensurate otherwise, the line $\Delta=1-h$
corresponding to a Pokrovski-Talapov (PT) phase transition
\cite{prokov} between these two regimes. Since the spectra
are equivalent, so are the phase diagrams.

To see the connection between $H_{0}^{\prime}$ and $H^{XXZ}$
we rewrite the latter in the basis of $E^{kl}$ matrices
\bea\label{eq:new}
H^{XXZ}&=&-\sum_{i=1}^{L-1}\left( E_{i}^{01}E_{i+1}^{10}+
E_{i}^{10}E_{i+1}^{01}\right) +vE_{i}^{11}E_{i+1}^{11}\nonumber\\
&&+ w\left( E_{i}^{11}+ E_{i+1}^{11} \right) +
a\left( E_{i}^{11}-E_{i+1}^{11} \right) + b
\eea
where we redefined the parameters of the Heisenberg
model as follows
\be\label{eq:heisenhub}
v=-2\Delta, \quad w=\Delta +h, \quad
b=-\bigl( \frac{\Delta +\beta}{2} + h \bigr)
\ee
$H_{0}^{\prime}$ is obtained from $H^{XXZ}$ by adding
extra terms proportional to $E^{02}_{i}E^{20}_{i+1}$
and $E^{20}_{i}E^{02}_{i+1}$
which do not affect the spectrum and correspond, in
the chemical scenario, to the exchange of the lattice
configuration $B\vac$ to $\vac B$ and $\vac B$ to $B\vac $
at sites $i$ and $i+1$ respectively. The diagonal terms are
extended accordingly.
The  wave functions of $H^{\prime}_{0}$ have been calculated
but the effect of $H_{1}^{\prime}$ on the phase diagram
has not yet been studied \cite{alc2}.
We shall reinterpret the known phase-diagram in the language
of chemical reactions.

The second class of chains which appear naturally
as time-evolution operators of chemical systems are the
$U_{q}SU(P/M)$-invariant Perk-Schultz models (PS models).
The general $N$-state Hamiltonian of the PS chains can be
written as
\bea\label{eq:supmq}
H_{\{\epsilon_{1},\cdots ,\epsilon_{P+M}\}}
&=&\sum_{j=1}^{L-1} U_{j}^{(P/M)}\nonumber\\
&=&\sum_{j=1}^{L-1}\biggl\{\frac{q+\qin}{2}-\biggl[
\sum_{\alpha\ne\beta =0}^{N-1}E_{j}^{\alpha\beta}
E_{j+1}^{\beta\alpha}
+\frac{q+\qin}{2}\sum_{\alpha =0}^{N-1}
\epsilon_{\alpha}
E_{j}^{\alpha\alpha}E_{j+1}^{\alpha\alpha} \nonumber\\
&&+\frac{q-\qin}{2} \sum_{\alpha\ne\beta=0}^{N-1}
sign\left(\alpha-\beta\right)
E_{j}^{\alpha\alpha}E_{j+1}^{\beta\beta}
\biggr]\biggr\}
\eea
with $\epsilon_{0}=\epsilon_{1}=\dots =\epsilon_{P-1}=
-\epsilon_{P}=-\epsilon_{P+1}=\dots = -\epsilon_{P+M-1}=1$.
Here $(P/M)$ stands for the entire partitions of $N$.
These models were first introduced by Sutherland \cite{sut}
for $q=1$ and then extended by Perk and Schultz
\cite{cher} to $q\ne 1$.
Later \cite{deg} it was noticed that
the $U_{j}^{(P/M)}$ are generators
of the Hecke algebra $\sl{H} (n)$ ($n=L-1$) defined through
\bea\label{eq:hecke}
U_{i} U_{i}&=&(q+\qin)U_{i}\qquad\qquad i=1,2,\cdots,n\nonumber\\
U_{i} U_{i\pm 1} U_{i} - U_{i}&=&U_{i\pm 1} U_{i} U_{i\pm 1}
-U_{i\pm 1}\nonumber\\
U_{i} U_{i+j}&=&U_{i+j}U_{i}\qquad\qquad j\ge i+2
\eea
where to each partition $(P/M)$ of $N$ corresponds an
additional set of relations beyond the ones given
above, defining the so-called quotients of the algebra.
{}From a mathematical point
of view, this underlying algebraic structure
has important consequences as regards the spectrum
of each chain \cite{mart}. From the physical one,
each quotient corresponds to systems with different
properties.
For $P+M=2$, one has two chains: the $(2/0)$ chain which
is the spin-$\frac{1}{2}$ $XXZ$ model and the
$(1/1)$ chain which has also been extensively studied \cite{haye}
and is believed to describe the coverage dependence on fugacity
for xenon adsorption on copper \cite{jau}. It has
a PT phase transition in its
phase diagram at $q+\qin =2$ between
a commensurate ordered phase and an incommensurate one with
oscillating correlation functions.
For $P+M=3$ the $(2/1)$ chain has received most of the
attention due to its relevance in
understanding Anderson's $t-J$ model \cite{anderson}.

We also found a very remarkable property of the mapping
of chemical systems given by combinations of rates
(\ref{eq:diff}) through (\ref{eq:pol}) onto the quantum
chains we presented. It turned out that due to the structure
of the mapping not all the parameters which we started with
(the different reaction rates)
are important in settling the phase structure of chemical
reactions. This will become clear in the next sections, were
we present our results with more detail.

The third important point is that for certain chemical models
it is necessary to define a nonlocal
similarity transformation in order to rewrite the Hamiltonian
describing them in the standard form of the PS
Hamiltonians. This kind of transformation was already known
to map the periodic spin-$\frac{1}{2}$ XXX with an $XY$
interaction term onto the XXZ model with general boundaries
proportional to the volume of the system \cite{jhh}.
The transformation we found generalizes it to higher-state
chains and also takes account of more general
Dzialoshinsky-Moriya-type interactions \cite{dzia} parametrized
by different variables.

This paper is organized as follows.
In the second section we introduce the
formalism of the master equation on lattices and its mapping onto
nearest neighbor quantum chains. The strategy we will adopt to
identify chemical reactions with quantum chains is explained.
Section $3$ is devoted to
the $U_{q} \widehat{SU(2)}$-invariant model, where it
will serve as a workbench for the application of the
ideas developed in the previous section.
In section $4$ we rewrite the PS models with
external fields as reaction-diffusion Hamiltonians,
constructing explicitly the similarity
transformation which map the chemical systems
onto them. In section $5$ we study this transformation and
generalize it in two directions: first to higher-state models
parametrized by different diffusion rates
and second showing that
diffusion-processes on periodic lattices
are mapped on chains with more general boundary conditions.
Finally in section $6$ we summon our results and
some questions we are still faced with.
\section{The Master Equation and Quantum Chains}
\hspace{\parindent}
The master equation governs the evolution of the probability
distribution of Markov processes. Due to its almost
universal range of applicability, it is one of the most
important equations of Statistical Mechanics. Here we shall apply
it in the context of chemical processes on a chain.
Consider a one-dimensional lattice with L sites and open boundaries.
At each site $j$ we define a variable $\beta_j$ which takes $N$
integer values ($0,1,2,...,N-1$). To each possible configuration
$\lbrace\beta\rbrace =\lbrace\beta_1,...,\beta_L\rbrace$
of the lattice realized at time $t$ we attach a
probability distribution $P(\lbrace\beta\rbrace,t)$
whose time evolution is given by the master equation
\bea\label{eq:me}
\frac{\partial P(\lbrace\beta\rbrace,t)}{\partial t}
&=&{\sum_{k=1}^{L-1}}
\biggl\{ -\Omega_{\beta_{k},\beta_{k+1}}
P(\beta_1,\ldots,\beta_{L};t)\nonumber\\
&+&{\sum_{l,m=0}^{L-1}}^{\prime}
\g^{\beta_{k} + l, \beta_{k+1} +m}_{\beta_{k},\beta_{k+1}}
P(\beta_{1}, \ldots,\beta_{k}+l,\beta_{k+1}+m,\ldots,
\beta_{L};t)\biggr\}
\eea
Here and henceforth the prime in a sum
over $l$ and $m$ indicates exclusion of the pair
$l=m=0$. The $\g^{a,b}_{c,d}$ are transition rates which equal
the probability that, in a unit time step and at any site $j$, a state
$(\beta_j,\beta_{j+1})=(a,b)$ changes to a state
$(\beta^{\prime}_{j},\beta^{\prime}_{j+1})=(c,d)$.
We will assume throughout this paper that the
transition rates depend only on links (nearest-neighbor
interaction) and are homogeneous, i.e. site independent.
The $\Omega_{a,b}$
are related to the probability that a state
$(a,b)$ will not change after a unit time step.
{}From conservation of probability we see that they satisfy
\be\label{eq:consprob}
\Omega_{a,b}={\sum_{r,s}}^{'}\g^{a,b}_{r,s}
\ee
With the help of these definitions and the matrices $E^{kl}$ defined in
the introduction we can rewrite the master equation (\ref{eq:me}) as
\be \label{eq:se}
\frac{\partial \ket{\psi}}{\partial t} = -H\ket{\psi}
\ee
if we identify $\ket{\psi}$ as the probability
$P(\lbrace\beta\rbrace,t)$ and $H$ as
\bea \label{eq:haminit}
H&=&\sum_{j=1}^{L-1} H_j = \sum_{j=1}^{L-1}(U_j-T_j) \nonumber \\
U_j&=&\sum_{a,b=0}^{N-1}
\Omega_{a,b}E_j^{aa} E_{j+1}^{bb}
\nonumber\\
T_j&=&{\sum_{a,b,c,d=0}^{N-1}}^{\prime}
\g^{a,b}_{c,d}E_j^{ca}
E_{j+1}^{db}
\eea

The key to the whole process of identifying chemical processes
with known quantum chains lies in finding the appropriate set
of rates so that we can recast $H$ as
\be
H=H_{0} + \sum_{j=1}^{L-1}(h_{i} + h_{i+1}
+g_{i}-g_{i+1}) + H_{1}
\ee
so that the first two terms on the r.h.s are equivalent to
a quantum chain plus external field and surface term, and the
spectrum of $H$ is independent of $H_{1}$. The surface term is an
extra degree of freedom that we have: since they do not alter
the bulk properties of the system, we can always define them.
We shall look only for those
chemical processes whose spectrum is equivalent to that of
some quantum chain, therefore guaranteeing the equality
of the phase diagram. However the wave functions
are not the same. {\sl A priori} one can define a similarity
transformation of the form $A(\lambda)$ such that
\be
A(\lambda)HA^{-1}(\lambda)=H_{0}+\sum_{j=1}^{L-1}(h_{i}+h_{i+1}
+g_{i}-g_{i+1}) + H_{1}^{\prime}(\lambda)
\ee
such that for some given value $\lambda=\lambda_{0}$ we
have $H_{1}^{\prime}(\lambda_{0})=0$. If such a transformation
were found, then the wave functions of the chemical problem
could be obtained from that of
the quantum chain. No solution has yet been found and the
problem remains open.
\section {The $U_{q}\widehat{SU(2)}$ model}
\hspace{\parindent}
With the formalism developed in the last section, we shall
now address the problem of finding the
set of chemical reactions associated to a given quantum
chain. Our ultimate goal is to use the phase diagram of
the chain to explain the chemistry of reaction-diffusion
processes. In what follows we
identify our particles with $A=1$, $B=2$
and inert state $=\vac$.

We consider a system in which
particles $A$ and
$B$ diffuse to the right and to the left symmetrically, with
rates equal to unity
\be
\left\{\begin{array}{clll}
A+\vac \rightleftharpoons\vac +A&rate&\g^{1,0}_{0,1}=\g^{0,1}_{1,0}= 1\\
\vspace{0.2cm}
B+\vac \rightleftharpoons\vac +B&&\g^{2,0}_{0,2}=\g^{0,2}_{2,0}=1
\end{array}
\right.
\ee
In addition to these processes, we also allow the particles
to react according to the following $24$ vacuum-driven rates:\\
\vspace{0.5cm}
$\bullet$ annihilation
\be
\left\{\begin{array}{clll}
A+A\rightarrow\vac +\vac && \g^{1,1}_{0,0}\\
\vspace{0.2cm}
B+B\rightarrow\vac +\vac && \g^{2,2}_{0,0}\\
\vspace{0.2cm}
A+B\rightarrow\vac +\vac && \g^{1,2}_{0,0}\\
\vspace{0.2cm}
B+A\rightarrow\vac +\vac && \g^{2,1}_{0,0}
\end{array}
\right.
\ee
$\bullet$ coagulation
\be
\left\{\begin{array}{clll}
A+A\rightarrow A +\vac && \g^{1,1}_{1,0}\\
\vspace{0.2cm}
A+A\rightarrow\vac +A && \g^{1,1}_{0,1}\\
\vspace{0.2cm}
B+B\rightarrow B +\vac && \g^{2,2}_{2,0}\\
\vspace{0.2cm}
B+B\rightarrow\vac +B && \g^{2,2}_{0,2}
\end{array}
\right.
\ee
$\bullet$ death
\be
\left\{\begin{array}{clll}
A+\vac\rightarrow\vac +\vac && \g^{1,0}_{0,0}\\
\vspace{0.2cm}
\vac +A\rightarrow\vac +\vac && \g^{0,1}_{0,0}\\
\vspace{0.2cm}
B+\vac\rightarrow\vac +\vac && \g^{2,0}_{0,0}\\
\vspace{0.2cm}
\vac +B\rightarrow\vac +\vac && \g^{0,2}_{0,0}
\end{array}
\right.
\ee
$\bullet$ polymerisation
\be
\left\{\begin{array}{clll}
A+A\rightarrow B+\vac && \g^{1,1}_{2,0}\\
\vspace{0.2cm}
A+A\rightarrow\vac +B && \g^{1,1}_{0,2}\\
\vspace{0.2cm}
B+B\rightarrow A+\vac && \g^{2,2}_{1,0}\\
\vspace{0.2cm}
B+B\rightarrow\vac +B && \g^{2,2}_{0,2}
\end{array}
\right.
\ee
$\bullet$ trapping
\be
\left\{\begin{array}{clll}
A+B\rightarrow A+\vac && \g^{1,2}_{1,0}\\
\vspace{0.2cm}
B+A\rightarrow\vac +A && \g^{2,1}_{0,1}\\
\vspace{0.2cm}
A+B\rightarrow B+\vac && \g^{1,2}_{2,0}\\
\vspace{0.2cm}
B+A\rightarrow\vac +B && \g^{2,1}_{0,2}\\
\vspace{0.2cm}
A+B\rightarrow\vac +A && \g^{1,2}_{0,1}\\
\vspace{0.2cm}
B+A\rightarrow A+\vac && \g^{2,1}_{1,0}\\
\vspace{0.2cm}
A+B\rightarrow\vac +B && \g^{1,2}_{0,2}\\
\vspace{0.2cm}
B+A\rightarrow B+\vac && \g^{2,1}_{2,0}
\end{array}
\right.
\ee
With these processes and the technique developed in section $2$
we obtain the following Hamiltonian
\bea\label{eq:study}
H&=& H_{0}+
H_{1}\nonumber\\
H_{0}&=&\sum_{i=0}^{L-1}\biggl\{
-\left( E_{i}^{01}E_{i+1}^{10}+E_{i}^{10}E_{i+1}^{01}
+E_{i}^{02}E_{i+1}^{20}+E_{i}^{20}E_{i+1}^{02}\right)\nonumber\\
&&+\biggl( \g^{0,1}_{0,0}+ 1\biggr)
E_{i}^{00}E_{i+1}^{11}+\biggl( \g^{0,2}_{0,0}+1\biggr)
E_{i}^{00}E_{i+1}^{22}\nonumber\\
&&+\biggl( \g^{1,0}_{0,0}+1\biggr)
E_{i}^{11}E_{i+1}^{00}+\biggl( \g^{2,0}_{0,0}+1\biggr)
E_{i}^{22}E_{i+1}^{00}\nonumber\\
&&+\biggl( \g^{1,2}_{0,0}+\g^{1,2}_{0,1}+\g^{1,2}_{1,0}+
\g^{1,2}_{0,2}+\g^{1,2}_{2,0}\biggr) E_{i}^{11}E_{i+1}^{22}
\nonumber\\
&&+\biggl( \g^{2,1}_{0,0}+\g^{2,1}_{1,0}+\g^{2,1}_{0,1}+
\g^{2,1}_{2,0}+\g^{2,1}_{0,2}\biggr) E_{i}^{22}
E_{i+1}^{11}\nonumber\\
&&+\biggl( \g^{1,1}_{0,0}+\g^{1,1}_{0,1}+ g^{1,1}_{1,0}+
\g^{1,1}_{0,2}+\g^{1,1}_{2,0}\biggr) E_{i}^{11}E_{i+1}^{11}\nonumber\\
&&+\biggl( \g^{2,2}_{0,0}+\g^{2,2}_{0,1}+\g^{2,2}_{1,0}+
\g^{2,2}_{0,2}+\g^{2,2}_{2,0}\biggr) E_{i}^{22}E_{i+1}^{22}
\biggr\}\nonumber\\
\vspace{0.2cm}
H_{1}&=&\sum_{i=1}^{L-1}\biggl\{
\g^{0,1}_{0,0}E_{i}^{00}E_{i+1}^{01}
+\g^{1,0}_{0,0}E_{i}^{01}E_{i+1}^{00}+\g^{0,2}_{0,0}
E_{i}^{00}E_{i+1}^{02}\nonumber\\
&&+\g^{2,0}_{0,0}E_{i}^{02}E_{i+1}^{00}+\g^{1,1}_{0,0}
E_{i}^{01}E_{i+1}^{01}+\g^{2,2}_{0,0}E_{i}^{02}
E_{i+1}^{02}\nonumber\\
&&+\g^{1,1}_{1,0}E_{i}^{11}E_{i+1}^{01}+\g^{1,1}_{0,1}
E_{i}^{01}E_{i+1}^{11}+\g^{2,2}_{2,0}E_{i}^{22}
E_{i+1}^{02}\nonumber\\
&&+\g^{2,2}_{0,2}E_{i}^{02}E_{i+1}^{22}+\g^{1,1}_{2,0}
E_{i}^{21}E_{i+1}^{01}+\g^{1,1}_{0,2}E_{i}^{01}
E_{i+1}^{12}\nonumber\\
&&+\g^{2,2}_{1,0}E_{i}^{12}E_{i+1}^{02}+\g^{2,2}_{0,1}
E_{i}^{02}E_{i+1}^{12}+\g^{1,2}_{0,0}E_{i}^{01}
E_{i+1}^{02}\nonumber\\
&&+\g^{2,1}_{0,0}E_{i}^{02}E_{i+1}^{01}+\g^{1,2}_{1,0}
E_{i}^{11}E_{i+1}^{02}+\g^{2,1}_{0,1}E_{i}^{02}
E_{i+1}^{11}\nonumber\\
&&+\g^{1,2}_{0,1}E_{i}^{01}E_{i+1}^{12}+\g^{2,1}_{1,0}
E_{i}^{12}E_{i+1}^{01}+\g^{1,2}_{0,2}E_{i}^{01}
E_{i+1}^{22}\nonumber\\
&&+\g^{2,1}_{2,0}E_{i}^{22}E_{i+1}^{01}+\g^{2,1}_{0,2}
E_{i}^{02}E_{i+1}^{21}+\g^{1,2}_{2,0}E_{i}^{21}
E_{i+1}^{02}\biggr\}
\eea
The spectrum of $H$ coincides with that of $H_{0}$.
We found this solution
numerically, by successive trials: we started out with a minimum
set of reactions, namely those corresponding to diffusion  and we
added vacuum-driven rates one by one
until we had a $2$-body Hamiltonian structure
resembling the structure of the quantum chain we were interested in. On
top of these, we added more processes until we reached the point were
the spectrum obtained was different.
This type of property
is characteristic of nonhermitian phenomena and has
an important physical implication:
since the spectrum does not depend on each parameter of
$H_{1}$ independently but only on their
combinations which appear in $H_{0}$,
our system has $8$ effective parameters, which are the $8$
sums inside parentheses of equation (\ref{eq:study})
\footnotemark[2]\footnotetext[2]{Due to conservation of
probability each diagonal element of a Hamiltonian
must equal the sum of the nondiagonal entries in that same column.
We can see this if we look at the definitions of rates in
section $2$ and realize that the conservation of the probabilities
implies in a relation among rates which can be written as
$\sum_{k\ne l}H_{kl} = H_{ll}$.}.
Yet, the problem of identifying this effective chemical
Hamiltonian with
the chain given by $H^{\prime}$ in (\ref{eq:ohio}) requires
the reduction to an even smaller set of parameters since
$H^{\prime}$ depends only on four.
The solution is to find
the proper way of recombining the eight effective parameters of the
chemical model into a final set of four. We will solve the problem in
steps.

First, to avoid carrying too heavy
a notation throughout the text we rename groups
of rates regarding
the same category of processes as follows ($i=1,2$)
\bea
A_{i}&=&\g^{i,i}_{0,0}
\qquad\qquad annihilation\nonumber\\
A_{1,2}^{\pm}&=&\g^{1,2}_{0,0}
\pm \g^{2,1}_{0,0}\qquad annihilation\nonumber\\
D_{i}^{\pm}&=&\g^{0,i}_{0,0} \pm \g^{i,0}_{0,0}\qquad death
\nonumber\\
C_{i}^{+}&=&\g^{i,i}_{i,0}+\g^{i,i}_{0,i}\qquad coagulation
\nonumber\\
P_{1}^{+}&=&\g^{1,1}_{0,2}+\g^{1,1}_{2,0}\qquad polymerisation
\nonumber\\
P_{2}^{+}&=&\g^{2,2}_{0,1}+\g^{2,2}_{1,0}\qquad polymerisation
\nonumber\\
T^{\pm}&=&\lbrack \g^{1,2}_{1,0}\pm \g^{2,1}_{0,1}\rbrack +
\lbrack \g^{1,2}_{0,1}\pm \g^{2,1}_{1,0}\rbrack \nonumber\\
&+&\lbrack \g^{1,2}_{2,0}\pm \g^{2,1}_{0,2}\rbrack +
\lbrack \g^{1,2}_{0,2}\pm \g^{2,1}_{2,0}\rbrack\qquad trapping
\eea
The problem is solved through the
rearrangement of the rates into the following variables
\bea\label{eq:identif}
v&=&A_{1}+C_{1}^{+}+P_{1}^{+}-D_{1}^{+}-2\nonumber\\
w&=&\frac{D_{1}^{+}+D_{2}^{+}}{4}+1\nonumber\\
g&=&\frac{D_{2}^{+}-D_{1}^{+}}{4}\nonumber\\
a&=&-\frac{D_{1}^{-}+D_{2}^{-}}{4}\nonumber\\
l&=&\frac{D_{2}^{-}-D_{1}^{-}}{4}
\eea
together with $3$ conditions on the rates
\bea\label{eq:hubrel}
2w+2g+v&=&
A_{2}+C_{2}^{+}+P_{2}^{+}\nonumber\\
4w+2v&=&
A_{12}^{+}+T^{+}\nonumber\\
4l&=&A_{12}^{-}+T^{-}
\eea
After some simple algebraic manipulation, we can
rewrite $H_{0}$ as a new $\widetilde{H}_{0}$
which reads
\bea\label{eq:iduknow}
\widetilde{H}_{0}&=&\sum_{i=1}^{L-1}-
\left( E_{i}^{01}E_{i+1}^{10}+
E_{i}^{10}E_{i+1}^{01}+E_{i}^{02}
E_{i+1}^{20}+E_{i}^{20}E_{i+1}^{02}
\right)\nonumber\\
&&+w(\varepsilon_{i}^{0} +\varepsilon_{i+1}^{0})
+g(\varepsilon_{i}^{z} +\varepsilon_{i+1}^{z})
+a(\varepsilon_{i}^{0} -\varepsilon_{i+1}^{0})\nonumber\\
&&+l(\varepsilon_{i}^{z} -\varepsilon_{i+1}^{z})
+v \varepsilon_{i}^{0} \varepsilon_{i+1}^{0}\nonumber\\
\varepsilon^{0}&=&E^{11}+E^{22} \qquad
\varepsilon^{z}= E^{11}-E^{22}
\eea
Comparing this expression with $H_{0}^{\prime}$
of eq. (\ref{eq:ohio}) we conclude that they
have the same phase diagram since
the term in $\widetilde{H}_{0}$ having the
parameter $l$ as coefficient is a surface
contribution. To interpret the
phase diagram of the chemical model in terms
of the phase diagram of the $XXZ$ model we
still have to recover the
$U_{q}\widehat{SU(2)}$-symmetric spectrum, which can
be done by requiring that
$g=l=0$. In terms of chemical rates this condition means
\bea\label{eq:morte}
D_{1}^{+}=D_{2}^{+}\nonumber\\
D_{1}^{-}=D_{2}^{-}
\eea
This amounts to saying that $A$ and $B$ are indistinguishable
as we can see by examining equations (\ref{eq:identif}),
(\ref{eq:hubrel}) and (\ref{eq:morte}).
It suffices now to identify
the parameters of
the Heisenberg chain with the rates of our chemical model.
We obtain
\bea
h&=&\frac{A_{1}+C_{1}^{+}+P_{1}^{+}}{2}\nonumber\\
\Delta&=&1+ \frac{D_{1}^{+}-
\left( A_{1}+C_{1}^{+}+P_{1}^{+}\right) }{2}\nonumber\\
a&=&-\frac{D_{1}^{-}}{2}\nonumber\\
\beta&=&-1-\frac{A_{1}+C_{1}^{+}+P_{1}^{+}+D_{1}^{+}}{2}
\eea

The analysis is straightforward.
Making $h=0$ implies that
no rates {\bf but} death survive. In this situation
\bea
\Delta =1+\frac{D_{1}^{+}}{2}\nonumber\\
\beta = -\Delta
\eea
By varying the death rate we go from a massive
regime to a massless one, that is the time evolution
for the concentration of particles has an exponential
or algebraic fall-off respectively. This can be understood
on physical grounds: being a `one-particle' process, death
happens irrespectively of any other processes ocurring in the system,
i.e. it is not diffusion-limited since any particle can die alone.
It therefore outruns the characteristic time scale set by diffusion
and brings about a quicker decay.
With an external field we have
\be
\Delta + h -1 = \frac{D_{1}^{+}}{2}
\ee
If we take death with a non-zero probability, then
from the equation above one sees that our system is massive
ferromagnetic. In Chemistry this means having a ground state with
no particles (we identify spin up with $\vac$)
By varying the rate of death we approach the line
on the phase diagram given by $\Delta +h=1$.
The system undergoes a PT transition
\cite{prokov} when the energy of the state with just one particle
equals that of the state with no particles and it becomes the ground
state. We have a level-crossing: since death is absent the system
can evolve to a final steady state where only one particle is left.
\section{The Perk Schultz chains}
\subsection{The $U_{q}SU(P/M)$-invariant chain}
\hspace{\parindent}
It is a well established fact that the Hamiltonian of the
$6$-vertex model is the time-evolution operator for the
two-state asymmetric diffusion process \cite{kandel}. On
the same ground we expect that the PS chains
will play the role of time-evolution operators
of higher-state asymmetric diffusion processes since they
are the Hamiltonians of higher-state ice models.
This picture is however far from complete and can be
extended to encompass
more general chemical systems if we reinterpret the
additional reactions on chemical Hamiltonians
as external fields in the PS chains they are mapped onto.
For the sake of
completeness, we present first the results without external
fields and then we proceed with the more general models.

Following \cite{alc1}, we first consider a system in
which particles $A$ and
$B$ diffuse and interchange positions on the lattice according to
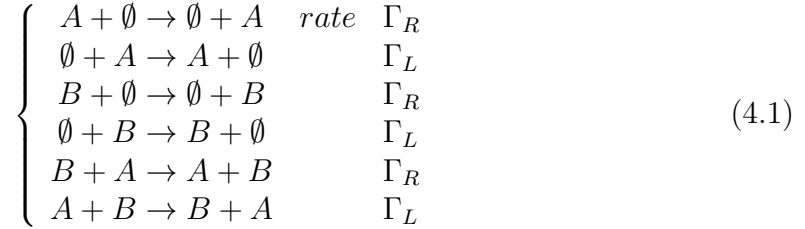
\be\label{eq:nef}
\left\{\begin{array}{clll}
A+\vac \rar\vac +A & rate & \Gamma_{R}\\
\vac+A\rar A+\vac && \Gamma_{L}\\
B+\vac \rar\vac +B && \Gamma_{R}\\
\vac+B\rar B+\vac && \Gamma_{L}\\
B+A\rar A+B && \Gamma_{R}\\
A+B\rar B+A && \Gamma_{L}
\end{array}
\right.
\ee
With these processes and the rates defined above we get
a Hamiltonian which reads
\bea
H&=&
\sum_{j=1}^{L-1}\biggl\{\g_{L}(E_{j}^{00}E_{j+1}^{11}+
E_{j}^{00} E_{j+1}^{22}+E_{j}^{11} E_{j+1}^{22})\nonumber\\
&&+\g_{R}(E_{j}^{11}E_{j+1}^{00}+E_{j}^{22}E_{j+1}^{00}+
E_{j}^{22}E_{j+1}^{11})\nonumber\\
&&-\g_{L} (E_{j}^{10}E_{j+1}^{01}+E_{j}^{20}E_{j+1}^{02}+
E_{j}^{21}E_{j+1}^{12})\nonumber\\
&&-\g_{R} (E_{j}^{01}E_{j+1}^{10}+E_{j}^{02}E_{j+1}^{20}+
E_{j}^{12}E_{j+1}^{21})
\biggr\}
\eea
Defining $\sqrt{\g_{L} \over \g_{R}}=q$,
which measures the asymmetry of the diffusion, and
the diffusion constant $\sqrt{\g_{L} \g_{R}}={\cal D}$,
which sets the time scale of the
problem, this can be rewritten as
\bea\label{eq:cheps1}
\frac{H_{\{\epsilon_{\alpha}\}}}{\cal{D}}&=&
\sum_{j=1}^{L-1}\biggl\{ \frac{q+\qin}{2}
-\biggl[ \qin\sum_{\alpha>\beta}^{2}
E_{j}^{\alpha\beta} E_{j+1}^{\beta\alpha}
+q\sum_{\alpha<\beta}^{2}
E_{j}^{\alpha\beta}
E_{j+1}^{\beta\alpha}\nonumber\\
&+&\frac{q+\qin}{2} \sum_{\alpha=0}^2
\epsilon_{\alpha}E_{j}^{\alpha\alpha}E_{j+1}^{\alpha\alpha} +
\frac{q-\qin}{2} \sum_{\alpha\ne\beta}^{2}
sign(\alpha-\beta)E_{j}^{\alpha\alpha}E_{j+1}^{\beta\beta}
\biggr]\biggr\}
\eea
where $(\epsilon_{0},\epsilon_{1},\epsilon_{2})=(1,1,1)$.
Comparing with eq. (\ref{eq:supmq}) we see that this is the
$U_{q}SU(3/0)$ chain in a non-standard form.
To bring it to
the standard form we have to consider a nonlocal similarity
transformation $S$ which is given by
\be\label{eq:similar}
S=\sum_{\alpha_{1},\alpha_{2},\cdots ,\alpha_{L} =0}^{2}
q^{\frac{1}{2} \sum_{j>i=1}^{L} sign\left(\alpha_{j}-
\alpha_{i} \right)}E^{\alpha_{1}\alpha_{1}}\otimes
E^{\alpha_{2}\alpha_{2}}\otimes\cdots\otimes E^{\alpha_{L}
\alpha_{L}}
\ee
The implications of this transformation coming
from boundary terms are discussed in detail
in the next section. If we apply it to the Hamiltonian
(\ref{eq:cheps1}) we obtain the standard $U_{q}SU(3/0)$
chain which reads
\bea\label{eq:ps}
S\frac{H_{\{1,1,1\}}}{\cal{D}}S^{-1}&=&
\frac{H^{(3/0)}}{\cal{D}}=
\sum_{j=1}^{L-1}\biggl\{ \frac{q+\qin}{2} -
\biggl[\sum_{\alpha\ne\beta}^{2}
E_{j}^{\alpha\beta} E_{j+1}^{\beta\alpha}\nonumber\\
&+&\frac{q+\qin}{2} \sum_{\alpha=0}^2
E_{j}^{\alpha\alpha}E_{j+1}^{\alpha\alpha}
+ \frac{q-\qin}{2}\sum_{\alpha\ne\beta}^{2}
sign(\alpha-\beta)E_{j}^{\alpha\alpha}E_{j+1}^{\beta\beta}
\biggr]\biggr\}
\eea
We use two notations to differentiate between what we call the
non-standard (non-hermitian) and standard (hermitian) forms of
the chain.

To reproduce the $U_{q}SU(2/1)$-invariant PS model,
two new processes have to be added, namely those
corresponding to coagulation of $B$
\be
\left\{\begin{array}{clll}
B+B\rar B+\vac & rate & \g^{2,2}_{2,0}\\
\vspace{0.2cm}
B+B\rar \vac+B & & \g^{2,2}_{0,2}
\end{array}
\right.
\ee
such that $\g^{2,2}_{2,0}+\g^{2,2}_{0,2}= {\cal D}( q+\qin )$.
The main difference now in comparison to the first case we studied
is that we will have an extra
diagonal contribution
equal to $\g^{2,2}_{2,0}+\g^{2,2}_{0,2}$ and two new nondiagonal
pieces proportional to each of these two rates. The Hamiltonian that
we find in this case reads
\be\label{eq:cheps2}
\frac{H}{\cal D}=
\frac{H_{\{1,1,-1\}}}{\cal{D}}-
\sum_{j=1}^{L-1}\biggl\{
\frac{\g^{2,2}_{2,0}}{\cal D}E_{j}^{22}E_{j+1}^{02}-
\frac{\g^{2,2}_{0,2}}{\cal D}E_{j}^{22}E_{j+1}^{20}\biggr\}
\ee
where $H_{\{1,1,-1\}}$ is obtained from (\ref{eq:cheps1})
with $(\epsilon_{0},\epsilon_{1},\epsilon_{2})=(1,1,-1)$.
The spectrum of $H$ is equal to the spectrum of
$H_{\{1,1,-1\}}$ and again we can study only the properties
of the latter as long as we are interested only in the phase
diagram. $H_{\{1,1,-1\}}$ is the non-standard representation
of the $(2/1)$ PS chain.
With the similarity transformation of eq. (\ref{eq:similar})
we obtain the standard
$U_{q}SU(2/1)$-invariant PS model
\be
SH_{\{1,1,-1\}}S^{-1}=H^{(2/1)}
\ee
where $H^{(2/1)}$ is given by (\ref{eq:supmq}) with
$(\epsilon_{0},\epsilon_{1},\epsilon_{2})=(1,1,-1)$.

Finally, the last PS chain can be reproduced
by the inclusion
of the rates corresponding to coagulation of $A$
\be
\left\{\begin{array}{clll}
A+A\rar A+\vac & rate & \g^{1,1}_{1,0}\\
\vspace{0.2cm}
A+A\rar \vac+A & & \g^{1,1}_{0,1}
\end{array}
\right.
\ee
with $\g^{11}_{10} + \g^{11}_{01}= {\cal D} (q+\qin )$.
The Hamiltonian in this case reads
\be\label{eq:cheps3}
\frac{H}{\cal D}=
\frac{H_{\{1,-1,-1\}}}{\cal{D}}-
\sum_{j=1}^{L-1}\biggl\{\frac{\g^{1,1}_{1,0}}{\cal D}
E_{j}^{11}E_{j+1}^{01}-
\frac{\g^{1,1}_{0,1}}{\cal D}E_{j}^{11}E_{j+1}^{10}\biggr\}
\ee
The same argument applies: the spectrum of $H$ is equal to
the spectrum of $H_{\{1,-1,-1\}}$, which is the non-standard
$U_{q}SU(1/2)$-invariant PS model. This is in turn
equivalent to the standard form through the trasformation $S$
\be
SH_{\{1,-1,-1\}}S^{-1}=H^{(1/2)}
\ee
We would like to point out one particular feature of the
two last chains.
Once we find the spectrum of the $(2/1)$ chain, the
spectrum of the $(1/2)$ can be obtained in a straighforward
manner: we reverse the sign of the whole spectrum of the $(2/1)$
model and an overall constant equal to $(L-1)(q+\qin)$, which is
the highest energy of the $(2/1)$ chain (this value is indeed
the highest energy for all $L$-site $(P/M)$ chains with $q$
real and non-zero $P$ and $M$.
An outline proof is given in ref. \cite{mart}).
In the chemical
scenario however, they correspond to different physical
pictures and the reason is the positivity of the spectrum.
\vskip 0.2in
\subsection{The $U_q SU(P/M)$-invariant
chain with external fields}
\hspace{\parindent}
We now generalize the results above to include
the most general set of vacuum-driven processes which
can be written as a pure nonhermitian
Hamiltonian which does not contribute to the spectrum of the
PS chains.
We analysed this problem on the computer and found that
beyond the $24$ rates of section $3$ we can include
one more set of reactions,
namely those corresponding to mutation and
transmutation of $A$. They are defined through
\vspace{0.5cm}
\be
\left\{\begin{array}{clll}
A+\vac \rar B+\vac & rate & \g^{1,0}_{2,0}\\
\vspace{0.2cm}
\vac +A\rar\vac +B & & \g^{0,1}_{0,2}\\
\vspace{0.2cm}
A+\vac\rar\vac +B & & \g^{1,0}_{0,2}\\
\vspace{0.2cm}
\vac +A\rar B+\vac & & \g^{0,1}_{2,0}
\end{array}
\right.
\ee
For these reactions we define the following combination
of rates
\bea
M_{1}^{\pm}&=&\g^{0,1}_{0,2}\pm \g^{1,0}_{2,0}\nonumber\\
X_{1}^{\pm}&=&\g^{0,1}_{2,0}\pm \g^{1,0}_{0,2}
\eea
The problem is the same of section $3$: the original
chemical process we start with has more parameters than the
PS chains. The question is  whether it is possible to reduce
these rates to the appropriate
set of parameters in a way that these
can be rewritten as external fields and surface terms.
We found a positive answer in all three cases studied.
Using the master equation approach we get a Hamiltonian which
can be written as two separate pieces as follows
\be\label{eq:ps1f}
\frac{H}{\cal{D}}=\frac{H_{\{\epsilon_{\alpha}\}}}{\cal D}
+\sum_{i=1}^{L-1}\left(
h_{i}+h_{i+1}+g_{i}-g_{i+1}\right) + H^{(1)}
\ee
where $H_{\{\epsilon_{\alpha}\}}$ is given by (\ref{eq:cheps1})
and $H^{(1)}=\sum_{i} H^{(1)}_{i}$ being the same
in all cases, reads
\bea\label{eq:t}
H^{(1)}_{i}&=&\g^{0,1}_{0,0}E_{i}^{00}E_{i+1}^{01}+\g^{1,0}_{0,0}E_{i}^{01}
E_{i+1}^{00}+\g^{0,2}_{0,0}E_{i}^{00}E_{i+1}^{02}\nonumber\\
&+&\g^{2,0}_{0,0}E_{i}^{02}E_{i+1}^{00}+\g^{1,1}_{0,0}E_{i}^{01}
E_{i+1}^{01}+\g^{2,2}_{0,0}E_{i}^{02}E_{i+1}^{02}\nonumber\\
&+&\g^{1,1}_{1,0}E_{i}^{11}E_{i+1}^{01}+\g^{1,1}_{0,1}E_{i}^{01}E_{i+1}^{11}
+\g^{2,2}_{2,0}E_{i}^{22}E_{i+1}^{02}\nonumber\\
&+&\g^{2,2}_{0,2}E_{i}^{02}E_{i+1}^{22}+\g^{1,1}_{2,0}E_{i}^{21}E_{i+1}^{01}
+\g^{1,1}_{0,2}E_{i}^{01}E_{i+1}^{12}\nonumber\\
&+&\g^{2,2}_{1,0}E_{i}^{12}E_{i+1}^{02}+\g^{2,2}_{0,1}E_{i}^{02}E_{i+1}^{12}
+\g^{1,2}_{0,0}E_{i}^{01}E_{i+1}^{02}\nonumber\\
&+&\g^{2,1}_{0,0}E_{i}^{02}E_{i+1}^{01}+\g^{1,2}_{1,0}E_{i}^{11}E_{i+1}^{02}
+\g^{2,1}_{0,1}E_{i}^{02}E_{i+1}^{11}\nonumber\\
&+&\g^{1,2}_{0,1}E_{i}^{01}E_{i+1}^{12}+\g^{2,1}_{1,0}E_{i}^{12}E_{i+1}^{01}
+\g^{1,2}_{0,2}E_{i}^{01}E_{i+1}^{22}\nonumber\\
&+&\g^{2,1}_{2,0}E_{i}^{22}E_{i+1}^{01}+\g^{2,1}_{0,2}E_{i}^{02}E_{i+1}^{21}
+\g^{1,2}_{2,0}E_{i}^{22}E_{i+1}^{02}\nonumber\\
&+&\g^{1,0}_{2,0}E_{i}^{21}E_{i+1}^{00}+\g^{0,1}_{0,2}E_{i}^{00}E_{i+1}^{21}
+\g^{1,0}_{0,2}E_{i}^{01}E_{i+1}^{20}\nonumber\\
&+&\g^{0,1}_{2,0}E_{i}^{20}E_{i+1}^{01}
\eea
The $h$'s and $g$'s are the external fields and surface terms.
Since they are different for each chain, we will present them
separately. Due to
the symmetry properties of the PS chains, the field
(and surface contribution) are decomposed into $3$ independent
variables $h_{\alpha}$ ($g_{\alpha})$,
one for each conserved particle of type $\alpha$.
Since one of our particles is an inert state,
we take the corresponding variables $h_{0}$ ($g_{0})$
equal to zero.
\vskip 0.2in
\noindent
{\bf i) The $U_q SU(3/0)$ chain}
\vskip 0.2in
For the first of our Hamiltonians the field and surface terms
are given by
\bea\label{eq:su3}
\frac{M_{1}^{+}+X_{1}^{+}+D_{1}^{+}}{2\cal{D}}&=&h_{1}\nonumber\\
\frac{M_{1}^{-}+X_{1}^{-}+D_{1}^{-}}{2\cal{D}}&=&-g_{1}\nonumber\\
\frac{D_{2}^{+}}{2\cal{D}}&=&h_{2}\nonumber\\
\frac{D_{2}^{-}}{2\cal{D}} &=& -g_{2}
\eea
together with the conditions
\bea\label{eq:relsu3}
M_{1}^{+}+X_{1}^{+}+D_{1}^{+}&=&C_{1}^{+}+P_{1}^{+}+A_{1}\nonumber\\
D_{2}^{+}&=&C_{2}^{+}+P_{2}^{+}+A_{2}\nonumber\\
A_{12}^{+}+T^{+}&=&M_{1}^{+}+X_{1}^{+}+D_{1}^{+}+D_{2}^{+}\nonumber\\
A_{12}^{-}+T^{-}&=&D_{2}^{-}-(M_{1}^{-}+X_{1}^{-}+D_{1}^{-})
\eea
\vskip 0.2in
\noindent
{\bf ii) The $U_q SU(2/1)$ chain}
\vskip 0.2in
In this case we get the following relations between field,
surface contributions and reaction rates
\bea\label{eq:su21}
\frac{M_{1}^{+}+X_{1}^{+}+D_{1}^{+}}{2\cal{D}}&=&h_{1}\nonumber\\
\frac{M_{1}^{-}+X_{1}^{-}+D_{1}^{-}}{2\cal{D}}&=&-g_{1}\nonumber\\
\frac{D_{2}^{+}}{2\cal{D}}&=&-h_{2}\nonumber\\
\frac{D_{2}^{-}}{2\cal{D}} &=&-g_{2}
\eea
For these identifications to hold we have to impose
the extra set of relations among rates
\bea\label{eq:relsu21}
M_{1}^{+}+T_{1}^{+}+D_{1}^{+}&=&C_{1}^{+}+P_{1}^{+}+A_{1}\nonumber\\
D_{2}^{+}&=&C_{2}^{+}+P_{2}^{+}+A_{2}-D(q+\qin)\nonumber\\
A_{12}^{+}+T^{+}&=&M_{1}^{+}+X_{1}^{+}+D_{1}^{+}+D_{2}^{+}\nonumber\\
A_{12}^{-}+T^{-}&=&D_{2}^{-}-(M_{1}^{-}+X_{1}^{-}+D_{1}^{-})
\eea
\vskip 0.2in
\noindent
{\bf iii) The $U_q SU(1/2)$ chain}
\vskip 0.2in
For the last model we have
\bea\label{eq:su12}
\frac{M_{1}^{+}+X_{1}^{+}+D_{1}^{+}}{2\cal{D}}&=&-h_{1}\nonumber\\
\frac{M_{1}^{-}+X_{1}^{-}+D_{1}^{-}}{2\cal{D}}&=&-g_{1}\nonumber\\
\frac{D_{2}^{+}}{2\cal{D}}&=&-h_{2}\nonumber\\
\frac{D_{2}^{-}}{2\cal{D}} &=&-g_{2}
\eea
The consistency conditions for these equations are
\bea\label{eq:relsu12}
M_{1}^{+}+T_{1}^{+}+D_{1}^{+}&=&C_{1}^{+}+P_{1}^{+}+A_{1}
-D(q+\qin )\nonumber\\
D_{2}^{+}&=&C_{2}^{+}+P_{2}^{+}+A_{2}-D(q+\qin)\nonumber\\
A_{12}^{+}+T^{+}&=&M_{1}^{+}+X_{1}^{+}+D_{1}^{+}+D_{2}^{+}\nonumber\\
A_{12}^{-}+T^{-}&=&D_{2}^{-}-(M_{1}^{-}+X_{1}^{-}+D_{1}^{-})
\eea
We observed that if in place of mutation and transmutation
of $A$ we considered the corresponding processes
for the particle $B$
\be
\left\{\begin{array}{clll}
\vac +B\rar\vac +A & rate & \g^{0,2}_{0,1}\\
\vspace{0.2cm}
B+\vac\rar\vac + A & & \g^{2,0}_{0,1}\\
\vspace{0.2cm}
\vac +B\rar A+\vac & &  \g^{0,2}_{1,0}\\
\vspace{0.2cm}
B+\vac\rar A+\vac & & \g^{2,0}_{1,0}
\end{array}
\right.
\ee
from which we define the relation of rates
\bea
M_{2}^{\pm}=\g^{0,2}_{0,1}\pm \g^{2,0}_{1,0}\nonumber\\
X_{2}^{\pm}=\g^{0,2}_{1,0}\pm \g^{2,0}_{0,1}
\eea
the spectrum also remained invariant, but not if we took
mutations and transmutations for both particles
at the same time.
The reason is that the inclusion of
both processes in the system yields a
local steady regime given by the reversible process
$A+\vac \rightleftharpoons B+\vac$.
In the presence of reversible reactions we expect
a totally different physical picture which translates
itself, among other things, into a different spectrum.
The phase diagram of the chemical processes can now
be in principle explained in terms of the Physics of
the PS models. Unfortunaltely only the phase
diagram in the absence of fields is known: this is however
not so interesting because it implies then that many processes
do not survive - and the whole mapping only makes sense if
one is able to study non-trivial cases. We will return to
this point in a future publication.
\section{Similarity Transformation
and Boundary Conditions}
\hspace{\parindent}
We shall now consider more thoroughly the transformation
given by the matrix $S$
which we used to rewrite our $3$-state Hamiltonians describing
diffusion processes as the standard
PS chains.
We shall address here two points: the first
regards the extention of $S$ not only to higher-state models
but also to multi-parameter diffusion processes, i.e. those
characterized by a larger set $\{ q \} = \{ q_{1},q_{2},
\cdots \}$ of parameters in substitution to the one-parameter
diffusion we dealt with so far. Second, we want to look at the effect
of $S$ on periodic chains and the physical features it induces.
\subsection{Multi-parameter and Higher-State Diffusion
Processes}
\hspace{\parindent}
As before we consider a
system in which $(N-1)$ particles diffuse with $\vac$ and
interchange places among themselves. We define the following rates
\be\label{eq:rager}
A_{x}+A_{y}\rightleftharpoons A_{y}+A_{x}
\;\;\;\; \Gamma^{A_{x},A_{y}}_{A_{y},A_{x}};\Gamma_{A_{x},A_{y}}^
{A_{y},A_{x}}
\;\;\;\; x>y=0,1,\cdots ,N-1
\ee
from which we also define $\frac{N(N-1)}{2}$ parameters $q_{xy}$
\be
\sqrt{\frac{\Gamma_{A_{x},A_{y}}^{A_{y},A_{x}}}{\Gamma_{A_{y},A_{x}}^
{A_{x},A_{y}}}}= q_{xy}
\ee
with $q_{xy}=q_{yx}^{-1}$. These are the generalization of
the parameter $q$ defined in the previous sections.
We impose a homogeneous time scale
\be
\sqrt{\Gamma_{A_{1},0}^{0,A_{1}}\Gamma_{0,A_{1}}^{A_{1},0}}=
\sqrt{\Gamma_{A_{2},0}^{0,A_{2}}\Gamma_{0,A_{2}}^{A_{2},0}}=\cdots =
\sqrt{\Gamma_{A_{N-1},A_{N-2}}^{A_{N-2},A_{N-1}}
\Gamma_{A_{N-2},A_{N-1}}^{A_{N-1},A_{N-2}}}=\cal{D}
\ee
and obtain a nonhermitian Hamiltonian which reads
\be\label{eq:geral}
\frac{H}{\cal D}=\sum_{k=1}^{L-1}\biggl\{ \sum_{x\ne y=0}^{N-1}
q_{xy}\left(E_{k}^{xx}E_{k+1}^{yy} - E_{k}^{xy}E_{k+1}^{yx}\right)
\biggr\}
\ee
We want to find a similarity transformation which makes $H$
hermitian. The (nonlocal!) similarity transformation
which accomplishes this reads
\be\label{eq:urtiga1}
S=\sum_{\alpha_{1},\alpha_{2},\cdots ,\alpha_{L}=0}^{N-1}
\biggl( \prod_{x>y=0}^{N-1}
(q_{xy})^{f_{xy}(\alpha_{1},\alpha_{2},\cdots ,\alpha_{L})}\biggr)
E^{\alpha_{1}\alpha_{1}}\otimes E^{\alpha_{2}\alpha_{2}}\otimes
\cdots \otimes E^{\alpha_{L}\alpha_{L}}
\ee
with the functions $f_{xy}$ given by
\be\label{eq:urtiga2}
f_{xy}(\alpha_{1},\alpha_{2},\cdots ,\alpha_{L})=\frac{1}{2}
\sum_{n>m=1}^{L}\left(\delta_{\alpha_{n},x}\delta_{\alpha_{m},y}
-\delta_{\alpha_{n},y}\delta_{\alpha_{m},x}\right)
\ee
We point out in the expression for the function that the
order of its arguments is very important since they index
the sites on the chain, that is
$f(\cdots,\al_{j},\al_{j+1},\cdots ) \ne f(\cdots,
\al_{j+1},\al_{j},\cdots )$.
\vskip 0.2in
\noindent
{\bf Proof}

For clarity, we will consider the diagonal and nondiagonal
pieces of the Hamiltonian separately. Since $S(\sum_{j}
H_{j,j+1})S^{-1} = \sum_{j} SH_{j,j+1} S^{-1}$, it suffices
to consider the action of $S$ on the $2$-site operator only.

Consider first the action of $S$ on the diagonal piece of the
Hamiltonian. We have
\bea
S\frac{H^{diag}_{j,j+1}}{\cal D}S^{-1}&=&
\sum_{\{\alpha\}=0}^{N-1}\prod_{x>y}\biggl(
q_{xy}^{f_{xy}\left(\alpha_{1},\alpha_{2},\cdots ,\alpha_{L}
\right)}
E^{\alpha_{1}\alpha_{1}}\otimes E^{\alpha_{2}\alpha_{2}}\otimes
\cdots \otimes E^{\alpha_{L}\alpha_{L}}\biggr)\nonumber\\
&\times&\biggl(\sum_{a\ne b=0}^{N-1}
q_{ab}E_{j}^{aa}E_{j+1}^{bb}\biggr)\nonumber\\
&\times&\sum_{\{\beta\}=0}^{N-1}
\prod_{w>z}\biggl(
q_{wz}^{-f_{wz}\left(\beta_{1},\beta_{2},\cdots ,\beta_{L}
\right)}
E^{\beta_{1}\beta_{1}}\otimes E^{\beta_{2}\beta_{2}}\otimes
\cdots \otimes E^{\beta_{L}\beta_{L}}\biggr)
\eea
We rearranje the terms in the expression above as follows
\bea
S\frac{H^{(diag)}_{j,j+1}}{\cal D}S^{-1}&=&\sum_{\{\al\}}
^{N-1}\sum_{\{\bt\}}^{N-1}\prod_{x>y=0}^{N-1}
\prod_{w>z=0}^{N-1}
\biggl(q_{xy}^{f_{xy}(\al_{1},\al_{2},\cdots ,\al_{L})}
q_{wz}^{-f_{wz}(\bt_{1},\bt_{2},\cdots ,\bt_{L})}\biggr)
\nonumber\\
&\times&E^{\al_{1}\al_{1}}E^{\bt_{1}\bt_{1}}\otimes E^{\al_{2}\al_{2}}
E^{\bt_{2}\bt_{2}}\otimes\cdots\otimes {\bf 1}^{(j)}\otimes
{\bf 1}^{(j+1)}\otimes\cdots\otimes E^{\al_{L}\al_{L}}
E^{\bt_{L}\bt_{L}}
\nonumber\\
&\times&\sum_{\stackrel{\al_{j},\al_{j+1}}{\bt_{j},\bt_{j+1}}}
\sum_{a\ne b=0}^{N-1}\biggl( q_{ab}E^{\al_{j}\al_{j}}E^{aa}
E^{\bt_{j}\bt_{j}}\otimes E^{\al_{j+1}\al_{j+1}}E^{bb}
E^{\bt_{j+1}\bt_{j+1}} \biggr)
\eea
Since $x,y,w$ and $z$ serve only to index the same set $\{ q\}$
of diffusion parameters, we can take $x=w$, $y=z$. Also,
the multiplication properties of the matrices $E^{\al \bt}$
gives us
\be
E^{pq}E^{rs}=\delta_{q,r} E^{ps}
\ee
We have therefore
\bea
S\frac{H^{(diag)}_{j,j+1}}{\cal D}S^{-1}&=&
\sum_{\{\al\}}
^{N-1}\sum_{\{\bt\}}^{N-1}\prod_{x>y=0}^{N-1}
\biggl(q_{xy}^{f_{xy}(\al_{1},\al_{2},\cdots ,\al_{L})
-f_{xy}(\bt_{1},\bt_{2},\cdots ,\bt_{L})}\biggr)
\nonumber\\
&\times&E^{\al_{1}\al_{1}}\delta_{\al_{1},\bt_{1}}\otimes E^{\al_{2}\al_{2}}
\delta_{\al_{2},\bt_{2}}\otimes\cdots\otimes {\bf 1}^{(j)}\otimes
{\bf 1}^{(j+1)}\otimes\cdots\otimes E^{\al_{L}\al_{L}}
\delta_{\al_{L},\bt_{L}}
\nonumber\\
&\times&\sum_{\stackrel{\al_{j},\al_{j+1}}{\bt_{j},\bt_{j+1}}}
\sum_{a\ne b=0}^{N-1}\biggl( q_{ab}E^{aa}_{j}
E^{bb}_{j+1}\delta_{\al_{j},a}\delta_{\al_{j+1},b}
\delta_{\al_{j},\bt_{j}}\delta_{\al_{j+1},\bt_{j+1}}\biggr)
\eea
Taking into account the Kronecker's deltas we have
\bea
S\frac{H^{(diag)}_{j,j+1}}{\cal D}S^{-1}&=&\sum_{\{\al\}}
^{N-1}\prod_{x>y=0}^{N-1}
q_{xy}^{f_{xy}(\al_{1},\al_{2},\cdots ,\al_{L})
-f_{xy}(\al_{1},\al_{2},\cdots ,\al_{L})}
\biggl(\sum_{a\ne b=0}^{N-1}q_{ab}E^{aa}_{j}
E^{bb}_{j+1}\biggr)\nonumber\\
&\times&E^{\al_{1}\al_{1}} \otimes E^{\al_{2}\al_{2}}
\otimes\cdots\otimes {\bf 1}^{(j)}\otimes
{\bf 1}^{(j+1)}\otimes\cdots\otimes E^{\al_{L}\al_{L}}
\eea
The exponent of $q_{xy}$ is clearly zero. Summing over $\{\al\}$
we have finally
\bea
S\frac{H^{(diag)}_{j,j+1}}{\cal D}S^{-1}&=&\sum_{a\ne b=0}
^{N-1}q_{ab}{\bf 1}\otimes {\bf 1}\otimes \cdots \otimes
E_{(j)}^{aa}\otimes E_{(j+1)}^{bb} \otimes\cdots\otimes {\bf 1}
\nonumber\\
&=&\frac{H^{(diag)}_{j,j+1}}{\cal D}
\eea
This concludes the proof that the diagonal elements of the Hamiltonian
are not changed by the transformation generated by $S$.

To see the effect of $S$ on the nondiagonal piece we proceed as before.
We consider only the $2$-body operator acting on sites
$(j,j+1)$ and we especialize to one given value of the pair $(a,b)$.
We have
\be
S\frac{H^{(nond)}_{j,j+1}}{\cal D}S^{-1}=S\biggl(
q_{ab} E_{j}^{ab}E_{j+1}^{ba} + q_{ab}^{-1}E_{j}^{ba}
E_{j+1}^{ab}\biggr)_{a>b}S^{-1}
\ee
As in the diagonal case after multiplying the $E^{\al\bt}$ matrices
on each site we obtain
\bea
S\frac{H^{(nond)}_{j,j+1}}{\cal D}S^{-1}
&=&\sum_{\{\al\}}
^{N-1}\sum_{\{\bt\}}^{N-1}\prod_{x>y=0}^{N-1}
\biggl(q_{xy}^{f_{xy}(\al_{1},\al_{2},\cdots ,\al_{L})-
f_{xy}(\bt_{1},\bt_{2},\cdots ,\bt_{L})}\biggr)
\nonumber\\
&\times&E^{\al_{1}\al_{1}}\delta_{\al_{1},\bt_{1}}\otimes
\cdots\otimes {\bf 1}^{(j)}\otimes
{\bf 1}^{(j+1)}\otimes\cdots\otimes E^{\al_{L}\al_{L}}
\delta_{\al_{L},\bt_{L}}
\nonumber\\
&\times&\sum_{\stackrel{\al_{j},\al_{j+1}}{\bt_{j},\bt_{j+1}}}
\biggl( q_{ab}E^{ab}_{j}
E^{ab}_{j+1}\delta_{\al_{j},a}\delta_{\bt_{j},b}
+q_{ab}^{-1}E^{ba}_{j}E^{ab}_{j+1}\delta_{\al_{j},b}\delta_{\bt_{j},a}
\biggr)_{a>b}\nonumber\\
&\times&\delta_{\al_{j+1},\bt_{j}}\delta_{\al_{j},\bt_{j+1}}
\eea
This reduces to
\bea\label{eq:acima}
S\frac{H^{(nond)}_{j,j+1}}{\cal D}S^{-1}&=&\sum_{\{\al\}}
^{N-1}\prod_{x>y=0}^{N-1}
q_{xy}^{f_{xy}(\al_{1},\cdots ,\al_{j},\al_{j+1},\cdots ,\al_{L})
-f_{xy}(\al_{1},\cdots ,\al_{j+1},\al_{j},\cdots ,\al_{L})}
\nonumber\\
&\times&E^{\al_{1}\al_{1}} \otimes E^{\al_{2}\al_{2}}
\otimes\cdots\otimes {\bf 1}^{(j)}\otimes
{\bf 1}^{(j+1)}\otimes\cdots\otimes E^{\al_{L}\al_{L}}
\nonumber\\
&\times&\biggl( q_{ab}E^{ab}_{j}
E^{ab}_{j+1}\delta_{\al_{j},a}\delta_{\al_{j+1},b}
+ q_{ab}^{-1}E^{ba}_{j}E^{ab}_{j+1}\delta_{\al_{j},b}
\delta_{\al_{j+1},a}\biggr)_{a>b}
\eea
We have to analyse the exponent of $q_{xy}$ now. We once again point
out that the order of the arguments of the functions $f$ is important.
With this in mind we have
\bea
f_{xy}(\al_{1},\cdots ,\al_{j},\al_{j+1},\cdots ,\al_{L})&-&
f_{xy}(\al_{1},\cdots ,\al_{j+1},\al_{j},\cdots ,\al_{L})=
\nonumber\\
{1\over 2}(\delta_{\al_{j+1},x}\delta_{\al_{j},y}-
\delta_{\al_{j+1},y}\delta_{\al_{j},x})&-&{1\over 2}
(\delta_{\al_{j},x}\delta_{\al_{j+1},y}-\delta_{\al_{j},y}
\delta_{\al_{j+1},x})\nonumber\\
&=&\delta_{\al_{j+1},x}\delta_{\al_{j},y}-
\delta_{\al_{j+1},y}\delta_{\al_{j},x}
\eea
Substituting this expression on (\ref{eq:acima}) we get
\bea
S\frac{H^{(nond)}_{j,j+1}}{\cal D}S^{-1}
&=&\sum_{\{\al\}}
^{N-1}\prod_{x>y=0}^{N-1}
q_{xy}^{\delta_{\al_{j+1},x}\delta_{\al_{j},y}-
\delta_{\al_{j+1},y}\delta_{\al_{j},x}}\nonumber\\
&\times&\biggl( q_{ab}E^{ab}_{j}
E^{ab}_{j+1}\delta_{\al_{j},a}\delta_{\al_{j+1},b}
+ q_{ab}^{-1}E^{ba}_{j}E^{ab}_{j+1}\delta_{\al_{j},b}
\delta_{\al_{j+1},a}\biggr)_{a>b}\nonumber\\
&\times&E^{\al_{1}\al_{1}} \otimes E^{\al_{2}\al_{2}}
\otimes\cdots\otimes {\bf 1}^{(j)}\otimes
{\bf 1}^{(j+1)}\otimes\cdots\otimes E^{\al_{L}\al_{L}}
\eea
Summing over $\{\alpha\}$
gives us an identity matrix on each site. The
deltas also `kill' the sum over $\al_{j},\al_{j+1}$ such
that we will be left with
\bea
S\frac{H^{(nond)}_{j,j+1}}{\cal D}S^{-1}&=&
\prod_{x>y=0}^{N-1}\biggr(q_{xy}^{\delta_{b,x}
\delta_{a,y}-\delta_{a,x}\delta_{b,y}}
q_{ab}E^{ab}_{j}E^{ba}_{j+1}+
q_{xy}^{\delta_{a,x}\delta_{b,y}-
\delta_{a,y}\delta_{b,x}}q_{ab}^{-1}
E^{ba}_{j}E^{ab}_{j+1}\biggl)_{a>b}\nonumber\\
&=&\biggl(q_{ab}^{-1}q_{ab}E^{ab}_{j}E^{ba}_{j+1}+
q_{ab}q_{ab}^{-1}E^{ba}_{j}E^{ab}_{j+1}\biggr)_{a>b}
\nonumber\\
&=&(E^{ab}_{j}E^{ba}_{j+1}+
E^{ba}_{j}E^{ab}_{j+1})_{a>b}
\eea
This concludes our proof.

To finish we would like to indicate how in the one-parameter
$N$-state diffusion process this expression simplifies.
The rates are defined according to
\be
A_{x}+A_{y}\rightleftharpoons A_{y}+A_{x}
\;\;\;\; \Gamma_{R},\Gamma_{L}
\;\;\;\; x>y=0,1,\cdots ,N-1
\ee
for which the $L$-site Hamiltonian reads
\bea\label{eq:seila}
\frac{H}{\cal D}&=&\sum_{k=1}^{L-1}\biggl\{\frac{q+\qin}{2}-
\biggl[ \qin\sum_{\alpha >\beta =0}^{N-1}E_{k}^{\alpha\beta}
E_{k+1}^{\beta\alpha} +
q\sum_{\alpha <\beta =0}^{N-1}E_{k}^{\alpha\beta}
E_{k+1}^{\beta\alpha} \nonumber\\
&&+\frac{q+\qin}{2}\sum_{\alpha =0}^{N-1}
E_{k}^{\alpha\alpha}E_{k+1}^{\alpha\alpha} +
\frac{q-\qin}{2}\sum_{\alpha\ne \beta =0}^{N-1} sign(\alpha - \beta)
E_{k}^{\alpha\alpha}E_{k+1}^{\beta\beta}\biggr]\biggr\}
\eea
This is the non-standard $U_{q}SU(N/0)$-invariant
PS model. To derive the standard form we consider
the matrix $S$ from eq. (\ref{eq:urtiga1}) with $q_{xy}=q$ for
all $x,y$. By observing that in this case we have
\be
f_{1}(\alpha_{1},\alpha_{2},\cdots ,\alpha_{L})+
\cdots +
f_{N-1}(\alpha_{1},\alpha_{2},\cdots ,\alpha_{L})=
\frac{1}{2}\sum_{n>m=1}^{L} sign(\alpha_{n}-\alpha_{m})
\ee
we get for $S$ the following expression
\be\label{eq:simp}
S=\sum_{\alpha_{1},\alpha_{2},\cdots ,\alpha_{L}=0}^{N-1}
q^{\frac{1}{2}\sum_{n>m=1}^{L} sign(\alpha_{n} -\alpha_{m})}
E^{\alpha_{1}\alpha_{1}}\otimes E^{\alpha_{2}\alpha_{2}}\otimes
\cdots \otimes E^{\alpha_{L}\alpha_{L}}
\ee
It is a simple algebraic exercise to bring the chain
(\ref{eq:seila}) to its standard form under the action
of $S$. Actually the result holds for any of the
$U_{q}SU(P/M)$-invariant PS chains with $P+M=N$.
\vskip 0.2in
\subsection{Periodic Boundary Conditions}
\hspace{\parindent}
Due to its nonlocality, the effect of $S$
on free or periodic chains
is different. In the first case it brings the original
non-hermitian chemical model to a hermitian quantum chain.
In the second case, it brings the non-hermitian chemical model to
a hermitian quantum chain but with general boundary terms
which imply in a generalized Dzialoshinsky-Moriya-type of interaction
in the bulk \cite{dzia}. This transformation was known for many
years for $2$-state systems \cite{jhh}.

To understand the effect of $S$ on periodic chains, it is appropriate
to start with the simplest system, namely a particle $A$ diffusing
to the left and right with rates
\be
A+\vac\rightleftharpoons \vac +A
\;\;\;\; \Gamma_{R},\Gamma_{L}
\ee
where $\cal{D}$ and $q$ are defined in the usual way
(see section $4$). The Hamiltonian
is given by
\bea
\frac{H}{\cal D}&=&\sum_{k=1}^{L}\biggl\{\frac{q+\qin}{2}-
\frac{q+\qin}{2}\sum_{\alpha =0}^{1}
E_{k}^{\alpha\alpha}E_{k+1}^{\alpha\alpha} -
\frac{q-\qin}{2}\sum_{\alpha\ne \beta =0}^{1} sign(\alpha - \beta)
E_{k}^{\alpha\alpha}E_{k+1}^{\beta\beta}\nonumber\\
&&-\qin E_{k}^{10}E_{k+1}^{01} -qE_{k}^{01}E_{k+1}^{10}\biggr\}
\eea
It is convenient to change to the more familiar basis of
Pauli matrices, by making the identification
\bea
E^{00}&=&\frac{ {\bf 1} + \si^{z} }{2} \nonumber\\
E^{11}&=&\frac{ {\bf 1} - \si^{z} }{2} \nonumber\\
E^{01}&=&\si^{+} \nonumber\\
E^{10}&=&\si^{-}
\eea
In this basis the Hamiltonian can be written as
\be\label{eq:malvado}
\frac{H}{\cal D}=-\frac{1}{2}\sum_{k=1}^{L}\biggl\{
2\qin \si_{k}^{+}\si_{k+1}^{-} +2q\si_{k}^{-}\si_{k+1}^{+} +
\frac{q+\qin}{2}\si_{k}^{z}\si_{k+1}^{z}
%\frac{q-\qin}{2}\left( \si_{k+1}^{z}- \si_{k}^{z}\right) -
-\frac{q-\qin}{4}\biggr\}
\ee
The constant term is irrelevant in our discussion and we
will drop it for the time being.
Applying eq. (\ref{eq:simp}) to the remaining terms in
the chain we get
\bea
S\frac{H}{\cal D} S^{-1}=\frac{\widetilde H}{\cal D}&=&
-\frac{1}{2}\sum_{k=1}^{L-1}\biggl\{
2 \si_{k}^{+}\si_{k+1}^{-} +2\si_{k}^{-}\si_{k+1}^{+} +
\frac{q+\qin}{2}\si_{k}^{z}\si_{k+1}^{z} \biggr\}\nonumber\\
&&-\frac{1}{2}\biggl\{
2 q^{L}\si_{L}^{+}\si_{1}^{-} +2q^{-L}\si_{L}^{-}
\si_{1}^{+} +
\frac{q+\qin}{2}\si_{L}^{z}\si_{1}^{z} \biggr\}
\eea
We can bring this under a common summation sign if we define
the boundaries as
\bea\label{eq:front}
\si_{L+1}^{z}&=&\si_{1}^{z}\nonumber\\
\si_{L+1}^{+}&=&q^{L}\si_{1}^{+}\nonumber\\
\si_{L+1}^{-}&=&q^{-L}\si_{1}^{-}
\eea
With this definition we have
\be
\frac{\widetilde H}{\cal D}=
-\frac{1}{2}\sum_{k=1}^{L}\bigg\{
2 \si_{k}^{+}\si_{k+1}^{-} +2\si_{k}^{-}\si_{k+1}^{+} +
\frac{q+\qin}{2}\si_{k}^{z}\si_{k+1}^{z} \biggr\}
\ee
Since q is real, we have
a boundary term which is proportional to the volume of the
system, changing the whole structure of the
problem. In other words, the dynamics of
diffusion processes in a chain with
open boundaries are different from that of a periodic
chain \cite{malte}.

For higher-state
diffusion models on periodic lattices we will get, after
applying $S$, chains which cannot be put in terms of
simple generalized boundaries. Rather we get a bulk
interaction. To see this we first note that the
whole effect of $S$ is to change the off-diagonal
terms of the $2$-body chemical Hamiltonian at the boundary,
namely $E_{L}^{ab}E_{1}^{ba}$ and
$E_{L}^{ba}E_{1}^{ab}$. In the $2$-state case there are only
two possibilities for the pair $(a,b)$:
$(1,0)$ and $(0,1)$. Since opposite pairs are related by
an inversion of the power of $q$, we have effectively one
result which is $q^{L}$. In higher-state cases, the
power depends on the particular $(a,b)$ chosen.
We present the results for the $3$-state
model and the for the general $N$-state case in what follows.
\vskip 0.2in
\noindent
{\bf i) $3$-state diffusion process}
\vskip 0.2in
Here we have $(a,b)=(0,1);(0,2);(1,2)$ and the reversed pairs.
The action of S yields
\bea
S\{ q^{-1}E_{1}^{01}E_{L}^{10} \}S^{-1}&=&
q^{-2}E_{1}^{01}\otimes M_{1}\otimes M_{1} \otimes
\cdots \otimes M_{1} \otimes E_{L}^{10}\nonumber\\
\vspace{0.4cm}
M_{1}&=&\left(\matrix{q&&\cr
&q&\cr
&&1\cr}\right)
\eea
for $(a,b)=(0,1)$. Exchanging $a,b$ we get the
same structure but with $q$
replaced by $\qin$ as expected. For the
pair $(a,b)=(1,2)$ we obtain
\bea
S\{ q^{-1}E_{1}^{12}E_{L}^{21} \}S^{-1}&=&
q^{-2} E_{1}^{12}\otimes M_{2}\otimes M_{2}
\otimes \cdots \otimes
M_{2} \otimes E_{L}^{21}\nonumber\\
\vspace{0.4cm}
M_{2}&=&\left(\matrix{1&&\cr
&q&\cr
&&q\cr}\right)
\eea
Finally for $(a,b)=(0,2)$ we have
\be
S\{ q^{-1}E_{1}^{02}E_{L}^{20} \}
S^{-1}=q^{-2} E_{1}^{02}\otimes M_{1}M_{2}\otimes
M_{1}M_{2}\otimes \cdots \otimes M_{1}M_{2}\otimes
E_{L}^{21}
\ee
We see now that in the case under study, the effect of
the transformation cannot be put in the boundary terms
in contradistinction to the
$2$-state problem - the interaction is `spread'
over the bulk. The structure is analogous in higher-state
models as it is shown next.
\vskip 0.2in
\noindent
{\bf ii) $N$-state diffusion process}
\vskip 0.2in
Assume that $(a,b)=(i,j)$ such that $i<j = 0,1,\cdots, N-1$.
We obtain in this case
\be\label{eq:ma}
S\{ q^{-1}E_{1}^{ij}E_{L}^{ji} \}S^{-1}=
q^{-2}E_{1}^{ij}\otimes M\otimes M \otimes
\cdots \otimes M \otimes E_{L}^{ji}
\ee
where $M$ is an $N\times N$ matrix which reads
\be
M=\left(\matrix{1&&&&&&&&&&\cr
&\ddots&&&&&&&&&\cr
&&1&&&&&&&&\cr
&&&q&&&&&&&\cr
&&&&q^{2}&&&&&&\cr
&&&&&\ddots&&&&&\cr
&&&&&&q^{2}&&&&\cr
&&&&&&&q&&&\cr
&&&&&&&&1&&\cr
&&&&&&&&&\ddots&\cr
&&&&&&&&&&1\cr}\right)
\ee
where $q$ appears at the i-th and j-th positions in the
diagonal.
All other $M_{kk}$ are equal to $1$ for $k<i$ and $k>j$ and
equal to $q^{2}$ in the case $i<k<j$.
The interaction is spread along the bulk as we can see from
eq. (\ref{eq:ma}).
\section{Conclusions}
\hspace{\parindent}
We must start this summary with one question: what is the purpose
of this whole technique? Does it bring any new information to what
is already known regarding reaction-diffusion processes?

Reaction-diffusion models are {\sl per se} a very rich
and fascinating field, from a physical and mathematical point of
view. The range of physical phenomena which they encompass is
extensive, from dispersive transport in amorphous silicon
\cite{oren} to polymerisation processes like thermal
soliton-antisoliton interaction of {\sl trans}-polycetylene
\cite{spouge}, to name a few. Experimental studies have also
been conducted on one-dimensional coagulation models \cite{kroon}.
This alone justifies their study and
any new method which might shed some light into the structure of
these problems is highly welcome.

We showed in this paper that a large class of chemical
processes can be understood in terms of the
$U_{q}SU(P/M)$-invariant Perk-Schultz models
and the $U_{q}\widehat{SU(2)}$-invariant model
with external fields. Since both are integrable, this opens a new
perspective, namely of employing analytical methods to calculate
physical quantities of interest exactly. We showed how the
phase diagram in the latter case completely settles the phase
diagram in the chemical model, the corresponding massless and
massive regimes yielding two types of time-dependence for the
decay of the particles' concentration.
We showed also
what processes correspond to a field in the language of quantum
chains and that not each and every processes governs the Physics
but combinations of them.

In the Perk-Schultz models little is known about the phase
diagram with external fields. Since the effect of a field
on the spectrum is trivial, the problem is a feasible one. However
the limitation comes mostly from hindrances in numerical
techniques. To calculate the phase diagram one needs data from
large lattice sizes in order to make good extrapolations. Here
the integrability of the model comes in handy. The Bethe Ansatz
Equations which allow for good spectral evaluations are not known
in the case of open chains. For periodic chains we must take care.
The chemical Hamiltonians associated to the Perk-Schultz models
are usually not in a standard form. We found a similarity
transformation which brings them to the standard form, and whose
properties were not yet discussed in the literature except for the
$2$-state processes. This transformation
induces terms of physical relevance in the bulk which
are generalizations of the Dzialoshinski-Moriya interaction,
thus confirming the fact that
the Physics of nonequilibrium problems depends on the
boundary conditions imposed \cite{malte}.
The three-state Perk-Schultz models with
these new interactions are
soluble through the Bethe Ansatz technique, which in this case
yields the following set of coupled nonlinear
equations \cite{dahmen}
\bea
e^{\gamma (N_{0}+N_{1})} \epsilon^{M_1}_{1} \biggl(
\frac{sinh(\lambda_{k}^{(0)}+\epsilon_{0}\gamma /2)}
{sinh(\lambda_{k}^{(0)}-\epsilon_{0}\gamma /2)}\biggr)^{N}&=&
\prod_{\al =1}^{M_0}
\frac{sinh(\lambda_{k}^{(0)}-\lambda_{\al}^{(0)}+\epsilon_{1}\gamma)}
{sinh(\lambda_{k}^{(0)}-\lambda_{\al}^{(0)}-\epsilon_{0}\gamma)}
\nonumber\\
&\times&\prod_{\al =1}^{M_1}
\frac{sinh(\lambda_{k}^{(0)}-\lambda_{\al}^{(1)}-
\epsilon_{1}\gamma /2)}
{sinh(\lambda_{k}^{(0)}-\lambda_{\al}^{(1)}+
\epsilon_{1}\gamma /2)}\nonumber\\
e^{\gamma (N_{1}+N_{2})} \epsilon^{M_2}_{2}\epsilon^{M_0}_{0}&=&
\prod_{\al =1}^{M_0}
\frac{sinh(\lambda_{\al}^{(0)}-\lambda_{l}^{(1)}+
\epsilon_{1}\gamma /2)}
{sinh(\lambda_{\al}^{(0)}-\lambda_{l}^{(1)}-
\epsilon_{1}\gamma /2)}\nonumber\\
&\times&\prod_{\al =1}^{M_1}
\frac{sinh(\lambda_{\al}^{(1)}-\lambda_{l}^{(1)}-
\epsilon_{2}\gamma)}
{sinh(\lambda_{\al}^{(1)}-\lambda_{l}^{(1)}+
\epsilon_{1}\gamma)}
\eea
where $\{\epsilon_{\al}\}$ are the parameters of
the Perk-Schultz chains, and $\gamma$ is related
to $q$ through $ q=exp(\gamma )$. The $N_{i}$'s correspond
to the particular number of particles of type $i$ in
each charge sector and the $M_{i}$'s equal
the number of roots of the set of coupled
equations. They are obtained from the $N_{i}$'s through
\be
M_{i} = L - ( N_{0} + N_{1} +\cdots + N_{i} )
\ee
where $L$ is the lattice size.
We shall analyse this problem in a future publication.
\vskip 0.2in
\noindent
{\bf Acknowledgements}

It is a pleasure to acknowledge V. Rittenberg and F. C. Alcaraz who
brought this problem to my attention and shared their insights with
me. My most sincere thanks go also to H. Hinrichsen and B.
Wehefritz who carefully read the original in its draft
form. Finally I would like to express my gratitude
to H.-M. Babujian and A. Berkovich for the valuable suggestions
on the final manuscript.


\begin{thebibliography}{11}
\bibitem{smo} M. V. Smoluchowski, Z. Phys. Chem. {\bf 92} (1917), 215
\bibitem{bramson} M. Bramson and D. Griffeath, Ann. Prob. {\bf 8}
(1980), 183
\bibitem{tour} D. C. Torney and H. M. McConnel, J. Phys. Chem. {\bf 87}
(1983), 1941
\bibitem{kkan} K. Kang, P. Meakin, J. H. Oh and S. Redner, J.
Phys. A {\bf 17} (1984), L60
\bibitem{racz} Z. Racz, Phys. Rev. Lett. {\bf 55} (1985), 1107
\bibitem{lush} A. A. Lushnikov, Phys. Lett. {\bf 120A} (1987), 135
\bibitem{lebowitz} M. Bramson and J. L. Lebowitz, Phys. Rev.
Lett. {\bf 61} (1988), 2397
\bibitem{kusovkov} V. Kusovkov and E. Kotomin, Rep. Prog. Phys. {\bf 51}
(1988), 2397
\bibitem{avra} D. ben-Avraham, M. A. Burschka and C. R. Doering,
J. Stat. Phys. {\bf 60} (1990), 695
\bibitem{priv} D. Toussaint and F. Wilczek, J. Chem. Phys. {\bf 78}
(1983),2642; K. Kang and S. Redner, Phys. Rev. Letters {\bf 52} (1984),
955;V. Privman, J. Stat. Phys. {\bf 72 3,4} (1993), 845
\bibitem{red} K. Kang and S. Redner, Phys. Rev. {\bf A30}
(1984), 2833
\bibitem{chop} D. Chopard, M. Droz, T. Karapiperis and Z. R\'acz,
Phys. Rev. {\bf E47,1} (1993), R40;
M. A. Rodriguez, G. Abramson, H. Wio and A. Bru,
Phys. Rev. {\bf E48,2} (1993), 829
\bibitem{spouge} J. L. Spouge, Phys. Rev. Lett. {\bf 60} (1988), 871
\bibitem{glauber} R. J. Glauber, J. Math. Phys. {\bf 4} (1963), 294
\bibitem{heims}S. P. Heims, J. Chem. Phys. {\bf 45} (1966), 370;
K. Kawasaki, Phys. Rev. {\bf 145} (1966), 224;
L. P. Kadanoff and J. Swift, Phys. Rev. {\bf 165}
(1968), 310
\bibitem{alc1} F. C. Alcaraz, M. Droz, M. Henkel and V. Rittenberg,
Ann. Phys. {\bf 230} (1994), 250
\bibitem{devega} H. J. De Vega, Int. J. Mod. Phys. A {\bf 4,10}
(1989), 2371
\bibitem{kandel} D. Kandel, E. Domany and B. Nienhuis, J. Phys.
A: Math Gen. {\bf 23} (1990), L755 and references therein.
\bibitem{baxter} R. J. Baxter in `Exactly Solved Models in
Statistical Mechanics', Academic Press, London, 1982
\bibitem{alc3} F. C. Alcaraz and V. Rittenberg, Phys. Lett. B
{\bf 314} (1993), 377
\bibitem{gwa} L.-H. Gwa ans H. Spohn, Phys. Rev. A {\bf 46,2} (1992),
844
\bibitem{burgers} J. M. Burgers in `The Nonlinear Diffusion Equation',
Riedel, Boston, 1974
\bibitem{alc2} F. C. Alcaraz, D. Arnaudon, V. Rittenberg and M.
Scheunert, `Hubbard-like Models in the Infinite Repulsion Limit
and Finite-Dimensional Representations of the Affine Algebra',
CERN Preprint CERN-TH.6935/93, to appear in `Int. J. Mod. Phys. B'
\bibitem{mc} J. D. Johnson and B. M. Mckoy, Phys. Rev {\bf A46}
(1972), 1613; M. Takahashi and M. Suzuki, Prog. Theor. Phys.
{\bf 48} (1972), 2187
\bibitem{prokov} V. L. Pokrovskii and A. L. Talapov, Sov. Phys. JETP
{\bf 51} (1980), 134
\bibitem{sut} B. Sutherland, Phys. Rev. {\bf B12} (1975), 3795
\bibitem{cher} J. H. H. Perk and C. L. Schultz in `Non-Linear
Integrable Systems, Classical Theory and Quantum Theory', ed. M.
Jimbo and T. Miwa, World Scientific, Singapore, 1981;
C. L. Schultz, Phys. Rev. Letters {\bf 46} (1981), 629,
Physica {\bf 122} (1983), 71
\bibitem{deg} T. Deguchi, J. Phys. Soc. Japan {\bf 58} (1989), 3441;
T. Deguchi and Y. Akutsu, J. Phys. A. {\bf 23} (1990), 1861
\bibitem{mart} P. P. Martin and V. Rittenberg, Int. J. Mod. Phys.
{\bf A7}, Suppl.1B (1992), 707
\bibitem{haye} H. Saleur in `Trieste Conference on Recent
Developments in Conformal Field Theories', World Scientific, 1989
\bibitem{jau} M. Jaubert, A. Glachand, M. Bienfat and G. Boato, Phys.
Rev. Letters {\bf 46} (1981), 1676
\bibitem{anderson} P. W. Anderson, Science {\bf 235} (1987), 1196;
P. A. Bares and G. Blatter, Phys. Rev. Letters {\bf 64}
(1990), 2567
\bibitem{jhh} J. H. H. Perk, private communication to V. Rittenberg.
\bibitem{dzia} I. Dzialoshinsky, J. Phys. Chem. Solids {\bf 4} (1958),
241; T. Moriya, Phys. Rev.{\bf 117} (1960), 635 and in `Magnetism', eds.
G. T. Rado and H. Suhl, Academic Press, London, 1963
\bibitem{malte} M. Henkel and G. Sch\"utz,
`Boundary-Induced Phase Transitions in Equilibrium and Non-Equilibrium
Problems', Univ. Gen\`eve preprint UGVA-DPT 07-826 (1993); F. C.
Alcaraz, `Exact Steady States of Asymmetric Diffusion and
Two-Species Annihilation with Back Reaction from the Ground State of
Quantum Spin Models', Universidade Federal de S\~ao Carlos - Brazil
preprint (1994)
\bibitem{oren} M. Hvam and M. H. Brodsky, Phys. Rev. Lett. {\bf 46}
(1981), 371; J. Orenstein and M. Kastner, Phys. Rev. Lett. {\bf 46}
(1981), 1421 and references therein
\bibitem{kroon} R. Kroon, H. Fleurent and R. Sprik, Phys. Rev. {\bf
E 47,4} (1993), 2462
\bibitem{dahmen} S. R. Dahmen, in preparation.
\end{thebibliography}
\end{document}